\documentclass[aps, prl, twocolumn, tightenlines, notitlepage, superscriptaddress, nofootinbib, preprintnumbers, floatfix, showkeys,10pt]{revtex4-2}

\usepackage[normalem]{ulem}
\usepackage{amssymb}
\usepackage{amsmath}
\usepackage{graphicx}
\usepackage{url}
\usepackage{ulem}
\usepackage[utf8]{inputenc}
\usepackage{slashed}
\usepackage{lipsum}
\usepackage[T1]{fontenc}
\usepackage{appendix}
\usepackage{csquotes}
\usepackage[x11names]{xcolor}

\usepackage[export]{adjustbox} 
\usepackage{textgreek} 
\usepackage{caption, subcaption} 
\usepackage{ragged2e} 
\DeclareCaptionJustification{justified}{\justifying}
\usepackage{booktabs} 
\usepackage{orcidlink}

\usepackage{hyperref}

\hypersetup{colorlinks,citecolor= nicered,linkcolor= blue}
\definecolor{nicered}{rgb}{0.7,0.1,0.1}
\definecolor{nicegreen}{rgb}{0.1,0.5,0.1}

\def\cevns{CE\textnu NS}
\def\d{\mathrm{d}}
\newcommand{\qtransfer}{\left|\mathbf{q}\right|}

\definecolor{byzantium}{rgb}{0.44, 0.16, 0.39}

\makeatletter
    \newcommand{\colorboxed}[3][white]{\fcolorbox{#2}{#1}{\m@th$\displaystyle#3$}}
\makeatother

\AtBeginDocument{\hypersetup{citecolor=byzantium,linkcolor=byzantium,urlcolor=byzantium}}
\begin{document}

\bibliographystyle{utphys}

\title{Axial-vector neutral-current measurements\texorpdfstring{\\}{} in coherent elastic neutrino-nucleus scattering experiments}%

\author{D. Aristizabal Sierra~\orcidlink{0000-0001-5429-3708}}%
\email{daristizabal@uliege.be}%
\affiliation{Universidad T\'ecnica Federico Santa Mar\'{i}a - Departamento de F\'{i}sica\\ Casilla 110-V, Avda. Espa\~na 1680, Valpara\'{i}so, Chile}%

\author{Pablo M. Candela~\orcidlink{0009-0009-8416-9295}}%
\email{pamuca@ific.uv.es}
\affiliation{Instituto de F\'{i}sica Corpuscular (CSIC-Universitat de Val\`{e}ncia), Parc Cient\'ific UV C/ Catedr\'atico Jos\'e Beltr\'an, 2 E-46980 Paterna (Valencia) - Spain}%

\author{Valentina De Romeri~\orcidlink{0000-0003-3585-7437}}%
\email{deromeri@ific.uv.es}
\affiliation{Instituto de F\'{i}sica Corpuscular (CSIC-Universitat de Val\`{e}ncia), Parc Cient\'ific UV C/ Catedr\'atico Jos\'e Beltr\'an, 2 E-46980 Paterna (Valencia) - Spain}

\author{Dimitrios K. Papoulias~\orcidlink{0000-0003-0453-8492}}%
\email{dimitrios.papoulias@uni-hamburg.de}
\affiliation{Institute of Experimental Physics, University of Hamburg, 22761, Hamburg, Germany}

\author{Laura Trincado S.~\orcidlink{}}%
\email{laura.trincado@usm.cl}
\affiliation{Universidad T\'ecnica Federico Santa Mar\'{i}a - Departamento de F\'{i}sica\\ Casilla 110-V, Avda. Espa\~na 1680, Valpara\'{i}so, Chile}

\begin{abstract}
Coherent elastic neutrino-nucleus scattering (\cevns) is predominantly governed by vector neutral-current interactions, with subleading contributions arising from the axial current in nuclei with non-zero ground-state spin. Experimentally, the extraction of axial-current contributions has been so far of little interest, mainly because of the challenges its measurement entail. In this work, we investigate the relative size of the vector and axial components for target materials currently employed by the neutrino and dark matter experimental communities. We identify fluorine-based compounds as the most promising targets for probing the axial-current event rate. Among them, octafluoropropane ($\text{C}_3\text{F}_8$) emerges as a particularly suitable candidate, given its widespread use in spin-dependent dark matter searches and its relevance for upcoming dedicated \cevns~experiments. Considering both pion decay-at-rest and reactor neutrino fluxes, we show that such measurements can allow an indirect determination of the axial coupling at the $\sim 10\%$ level, depending on flux uncertainties and detector thresholds. We further emphasize that measurements of the axial current will allow to probe spin-dependent new physics scenarios through \cevns.
\end{abstract}

\maketitle

\textit{Introduction}---Since its first observation in 2017~\cite{COHERENT:2017ipa}, coherent elastic neutrino-nucleus scattering (\cevns)~\cite{Freedman:1973yd} has enabled low-energy measurements of Standard Model (SM) parameters and provided a sensitive probe of new physics in the neutrino sector (see, e.g., Refs. \cite{DeRomeri:2022twg,Cadeddu:2020lky}). A well-established experimental program employing a variety of target materials and detector technologies is now underway. Measurements of \cevns~have been performed by the COHERENT collaboration using CsI, LAr and Ge targets~\cite{COHERENT:2017ipa,COHERENT:2020iec,COHERENT:2021xmm,COHERENT:2024axu}, and more recently using reactor neutrinos with germanium detectors~\cite{Colaresi:2022obx,Ackermann:2025obx}. 
In parallel, the dark matter (DM) direct detection experiments XENONnT~\cite{XENON:2024ijk}, PandaX-4T~\cite{PandaX:2024muv} and LZ~\cite{LZ:2025igz} have as well reported on their first \cevns~observations on xenon targets. Although measurements using $^8$B solar neutrinos are currently statistically limited, rapid experimental progress is expected to significantly improve sensitivities and establish these facilities as key contributors to the global \cevns~research program. Additional measurements using alternative targets and novel detection techniques are either underway or planned~\cite{Abdullah:2022zue}. All these efforts are pushing \cevns~toward a precision era characterized by reduced experimental and theoretical uncertainties and substantially larger data samples.

In this emerging precision regime, subleading effects in the scattering process become increasingly relevant~\cite{AtzoriCorona:2024rtv,Tomalak:2020zfh,Mishra:2023jlq}.  Among these, the axial neutral-current contribution to the total event rate is of particular importance. To date, this contribution has received comparatively little attention, mainly because of the properties of the detector targets employed so far. Because the axial interaction depends on nuclear spin, its contribution is strongly suppressed relative to the vector component, which instead encodes the typical coherent enhancement of the process. This suppression is especially pronounced in heavy nuclei~\cite{Barranco:2005yy}, which are those that have been predominantly used in measurements so far. 
An important exception is $^{40}$Ar, an even–even nucleus with zero ground-state spin, for which no axial contribution is expected due to angular momentum conservation. Upcoming experiments with larger fiducial masses and a broader range of target materials are expected to change this picture, opening the possibility of experimentally accessing axial-current effects in \cevns. 

Although the general qualitative trend is well understood, a quantitative assessment of the most suitable target materials and the corresponding expected event rates is still required.
Access to the axial contribution is essential for a complete understanding of low-energy neutral current neutrino nucleus scattering physics. Its measurement would open new avenues for spin-dependent new physics probes, an aspect that has not yet been systematically explored in this context. Furthermore, just as the vector component has provided valuable information on SM parameters, sensitivity to the axial current would provide indirect information on the axial-vector coupling $g_A$, traditionally measured in cold neutron experiments (see, e.g., \cite{Markisch:2018ndu}), and, depending on achievable statistics and detector performance, potentially also on the strange axial-vector coupling $g_A^s$.

Motivated by these considerations, in this work we study axial-current event rates by comparing their magnitude with that of the vector contribution. Guidance comes from the extensive program of direct-detection experiments~\cite{MarrodanUndagoitia:2015veg,Schumann:2019eaa,Billard:2021uyg} searching for spin-dependent DM interactions, where nuclei with non-zero spin play a central role. In particular, the nuclear spin structure of the relatively light nucleus $^{19}$F has long been recognized as especially favorable for probing axial interactions due to its large and well-understood spin expectation values~\cite{PICO:2017tgi}.  
Building on this insight, we identify fluorine-based compounds as the most suitable target materials for accessing the axial contribution in \cevns~measurements. In particular, among carbon tetrafluoride ($\text{C}\text{F}_4$), octafluoropropane ($\text{C}_3\text{F}_8$) and perfluorobutane ($\text{C}_4\text{F}_{10}$)---which have been successfully tested by PICASSO~\cite{Archambault:2009sm,PICASSO:2012ngj} and PICO~\cite{PICO:2017tgi,PICO:2019vsc} in DM
searches---we find $\text{C}_3\text{F}_8$ to be the most favorable option due to its widespread use in the DM community and its demonstrated scalability to large detector masses. Bubble-chamber experiments such as PICO-60 have successfully operated detectors filled with 52 kg of $\text{C}_3\text{F}_8$, achieving leading sensitivity to spin-dependent interactions~\cite{PICO:2017tgi,PICO:2019vsc}. Further, discussions have been held for upgrades toward fiducial masses of up to 500 kg~\cite{Garcia-Viltres:2021swf}. 

In addition, experimental groups at the European Spallation Source (ESS)~\cite{Baxter:2019mcx} and the Japan Proton Accelerator Research Complex (J-PARC)~\cite{Collar:2025sle} have identified this compound as a feasible target for \cevns~measurements. Leveraging their insensitivity to electron recoil backgrounds~\cite{PICO:2019rsv}, moderately superheated bubble chambers with the integration of scintillation signals would facilitate sub-keV nuclear recoil thresholds \cite{Baxter:2019mcx,Baxter:2017ozv}.

Motivated by these prospects, in this letter we demonstrate that, under realistic detector performance a
statistically meaningful extraction of the axial current component as well as of the axial-vector coupling is possible. The achievable precision is primarily determined by detector threshold, fiducial mass and neutrino flux uncertainties.

\textit{Heavy versus medium-light targets axial-current contributions}---Phenomenological analyses of \cevns~data have so far primarily relied on the vector-current contribution to the differential cross section, as event rates are largely dominated by this component. A relevant quantity for this contribution is the weak-charge form factor, which quantifies the strength of the $Z$-boson coupling to nuclei. Usually, this form factor is typically described using phenomenological parametrizations, most commonly the Helm~\cite{Helm:1956zz} or the Klein-Nystrand~\cite{Klein:1999qj} models. The theoretical uncertainty implied by these choices is then incorporated as part of the systematic uncertainty in statistical analyses of the data. 
An alternative, though more fundamental, approach is based on \textit{ab initio} calculations of the weak-charge form factor~\cite{Papoulias:2019lfi,Hoferichter:2020osn}. In this approach, nucleon-nucleon interactions derived from chiral effective field theory are employed to construct the nuclear many-body Schr\"odinger equation. The resulting many-body wave functions are subsequently used to compute the form factor from first principles. Since both approaches carry intrinsic theoretical uncertainties, phenomenological parametrizations are generally preferred in actual data analyses due to their simplicity and ease of implementation.

For the axial current this is no longer the case\footnote{An approximate parametrization of the axial-vector form factor has been proposed in Refs. \cite{Belanger:2008sj,Cerdeno:2012ix}. At large transferred momentum, however, it provides a less accurate description than the result following from ab-initio calculations.}, and \textit{ab initio} results for the axial form factor become essential. The reason is that, in contrast to the vector contribution, the axial scattering rate exhibits an intrinsic dependence on nuclear spin and therefore on detailed nuclear-structure effects. The axial form factor is determined by the response of the nucleus to an external axial-vector current~\cite{DelNobile:2021wmp}. The different nuclear responses arise from a multipole decomposition of the hadronic current operator.
The dominant contributions are associated with the transverse spin current carried by the nucleons and are, following Ref. \cite{Hoferichter:2020osn}, denoted by $\mathcal{F}^{\Sigma_L^\prime}_{\pm}$. These quantities are provided in terms of fit functions obtained within shell-model calculations employing a harmonic-oscillator basis.

The axial form factor is obtained as a combination of isoscalar ($\mathcal{F}^{\Sigma_L^\prime}_+$) and isovector ($\mathcal{F}^{\Sigma_L^\prime}_-$) contributions (see Refs.~\cite{Hoferichter:2020osn,Klos:2013rwa} for details),
\begin{equation}
    \label{eq:Axial_FF}
    F_A = \frac{8\pi}{2J + 1}
    \left[(g_A^s)^2 S_{00}^\mathcal{T}
    - g_A\,g_A^s S_{01}^\mathcal{T}
    + g_A^2 S_{11}^\mathcal{T}
    \right]\ ,
\end{equation}
where $g_A^s$ quantifies the contribution of strange quarks to the spin structure of the nucleon and $J$ refers to the spin of the nuclear ground state. In writing Eq.~(\ref{eq:Axial_FF}), isospin symmetry has
been assumed. Restricting the calculation to leading-order spin-dependent one-body currents, the structure functions $S_{ij}^\mathcal{T}$ are given by~\cite{Hoferichter:2020osn}
\begin{align}
    \label{eq:structure_factors_S_ij}
    S_{00}^\mathcal{T}&= \sum_L[F_+^{\Sigma^\prime_L}]^2\ ,
    \nonumber\\
    S_{01}^\mathcal{T}&=\sum_L
    [F_-^{\Sigma^\prime_L}]^2\ ,
    \nonumber\\
    S_{11}^\mathcal{T}&=2\sum_L
    F_+^{\Sigma^\prime_L}F_-^{\Sigma^\prime_L}\ .
\end{align}

Here the sum runs over the contributing multipoles $L$. Although results including sub-leading pion-exchange two-body currents are available for several target nuclei, we do not include them in the present analysis. Their impact is expected to be subdominant compared to the experimental uncertainties associated with the extraction of the axial-current contribution itself. This constitutes a reasonable assumption, particularly  for a first generation of measurements aimed at probing the axial component in \cevns.

The differential cross section, including the axial-vector contribution, is then given by the standard vector term and the pure axial term, namely~\cite{Hoferichter:2020osn}
\begin{align}
    \label{eq:full_diff_cross_section}
    \frac{\d\sigma}{\d E_r}=&\frac{G_F^2  m_\mathcal{N}}{4\pi}
    \left(1 - \frac{m_\mathcal{N} E_r}{2 E_\nu^2} - \frac{E_r}{E_\nu}\right)
    Q_W^2 |F_W(\qtransfer^2)|^2
    \nonumber\\
    &+ \frac{G_F^2 m_\mathcal{N}}{4\pi}
    \left(1 + \frac{m_\mathcal{N} E_r}{2 E_\nu^2} - \frac{E_r}{E_\nu}\right)F_A(\qtransfer^2)\ .
\end{align}
Here $Q_W = 2(g_V^p Z + g_V^n N) = Z(1 - 4\,s_W^2) - N$, 
with $g_V^p = 1/2 - 2 s_W^2$, $g_V^n = -1/2$, and 
$s_W^2 \equiv \sin^2\theta_W$ denoting the weak-mixing angle. 
The quantity $F_W(\qtransfer^2)$ represents the weak-charge form factor,  for which we adopt the Helm parametrization~\cite{Helm:1956zz}. 
In principle, the simultaneous presence of vector and axial currents implies interference terms. These contributions, however, either vanish or are suppressed by recoil-energy factors~\cite{Hoferichter:2020osn}. 
As a result, to a very good approximation the vector and axial components contribute independently to the total scattering rate.

We will determine event rates using neutrino fluxes from  $\pi$ decay-at-rest at spallation neutron sources, and nuclear reactors. We assume an ESS-like spallation source, in which case, 0.3 neutrinos of each flavor are produced per proton-on-target (POT), assuming a 2 GeV proton-collision energy~\cite{Baxter:2019mcx}.
The reactor electron anti-neutrino flux is instead implemented according to Ref.~\cite{CONNIE:2019xid}. The relevant features of both sources are summarized in the Supplemental Material.
\begin{figure*}[t!]
    \centering
    \includegraphics[scale=0.6]{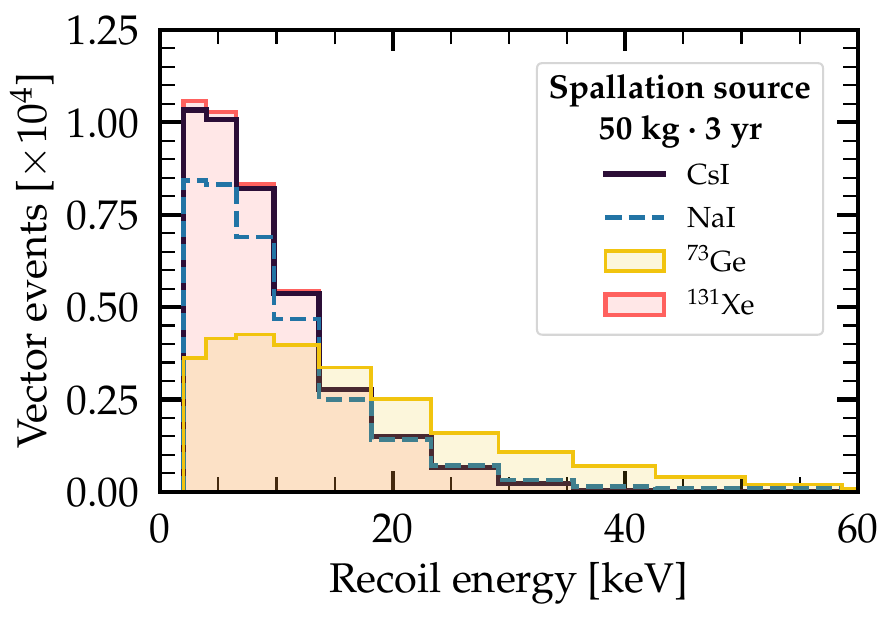}
    \hfill
    \includegraphics[scale=0.6]{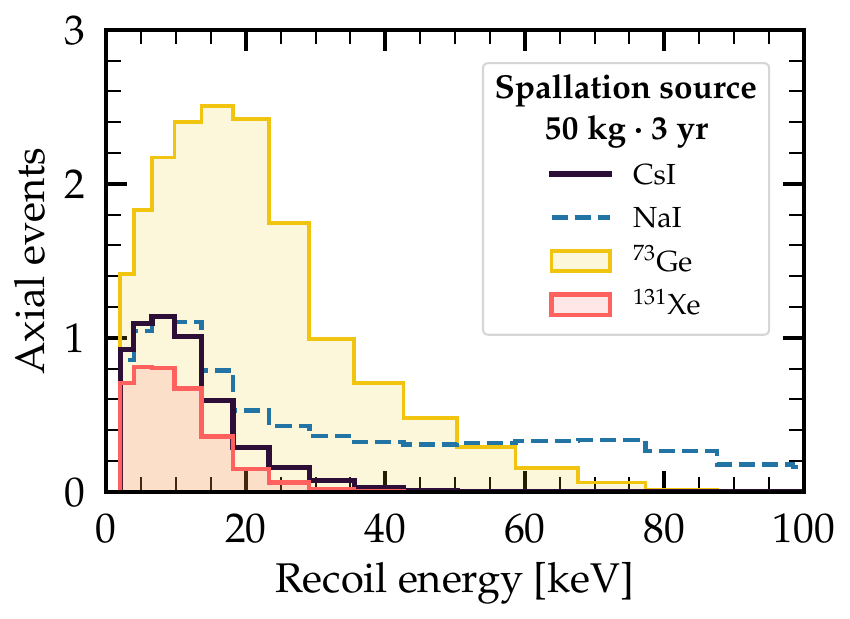}
    \caption{Expected vector event rates in heavy target materials (\textbf{left panel}) and the corresponding axial-current contributions (\textbf{right panel}), see the text for details. 
    }
    \label{fig:event_rates_heavy_targets}
\end{figure*}

Before turning to light compounds, we first evaluate the axial-current contribution for heavy target materials foreseen or already deployed at relevant \cevns~facilities. The COHERENT $\text{Na}\text{I}$ is primarily designed for charged-current measurements \cite{COHERENT:2023ffx}, but its technology can be scaled to tonne-size masses. In such a configuration, measurements in the nuclear recoil channel would, in principle, allow the
collection of either vector- and axial-current events. Both isotopes are odd-$A$, proton-odd nuclei with positive parity and non-zero spin: $J^\Pi(^{23}\text{Na})=3/2^+$ and $J^\Pi(^{127}\text{I})=5/2^+$.
Likewise, in CsI (already used by the COHERENT Collaboration~\cite{COHERENT:2017ipa,COHERENT:2021xmm}, and planned to be used at ESS~\cite{Baxter:2019mcx}, J-PARC~\cite{Collar:2025sle}, and the Chinese Spallation Neutron Source~\cite{CSNS,Su:2023klh}), the $^{133}\text{Cs}$ isotope is an odd-$A$, proton-dominated, positive-parity nucleus with $J^\Pi(^{133}\text{Cs})=7/2^+$. Therefore, in these compounds both nuclei contribute to the axial current.

The situation is different for xenon and germanium\footnote{Used by XENONnT~\cite{XENON:2024ijk}, LUX-ZEPLIN~\cite{LZ:2025igz} and PandaX-4T \cite{PandaX:2024muv}. In use as well by reactor-based \cevns~experiments \cite{Colaresi:2022obx,Ackermann:2025obx}.}. Among their stable isotopes, only $^{129}\text{Xe}$, $^{131}\text{Xe}$ and $^{73}\text{Ge}$ have non-zero spin ground states and can therefore generate axial-current events. Nevertheless, xenon and germanium remain among the most used detector materials, and highly enriched germanium detectors have been constructed for neutrino-less double $\beta$ decay searches~\cite{GERDA:2020xhi,Majorana:2013cem}. If the axial contribution is sufficiently large, one could envision dedicated enriched xenon or germanium detectors in which the axial signal is enhanced through isotopic purification.

We have calculated the expected event rates for these compounds assuming the following ESS-inspired~\cite{Baxter:2019mcx} benchmark configuration: $L=20\,\text{m}$ baseline, a proton beam power of $1.3$ MW, an $80\%$ flat detection acceptance with a $2$ keV recoil threshold, and a $50$ kg fiducial mass. Results are shown in Fig.~\ref{fig:event_rates_heavy_targets} for an exposure corresponding to three years of data taking. The left (right) panel shows the vector (axial) contribution.
In all cases, the axial contribution is strongly suppressed, with a clear trend such that the suppression increases with nuclear mass. As an order-of-magnitude estimate the relative size of the two contributions can be calculated from the weak-charge and axial form factors at zero transferred momentum
\begin{equation}
    \label{eq:order_of_mag_estimate}
    \frac{\text{Axial events}}{\text{Vector events}} \lesssim \frac{32 \pi}{3}\frac{J+1}{J (2J+1)}\frac{g_A^2 \left( \mathbb{S}_p-\mathbb{S}_n \right)^2 }{N^2}  \ .
\end{equation}
Here, $\mathbb{S}_p$ ($\mathbb{S}_n$) is the proton (neutron) spin expectation value and we have used the fact that the axial-current cross section is dominated by the isovector contribution, normalized according to $\mathcal{S}_{11}^\mathcal{T}(0)= \frac{4}{3}\frac{J+1}{J}g_A^2 \left(\mathbb{S}_p - \mathbb{S}_n \right)^2$~\cite{Hoferichter:2020osn}. 
The predicted event yields for both components are summarized in
Tab.~\ref{tab:heavy_compunds_coh_vs_axial}. Note that in our calculation we assume
a nuclear weak-charge radius
$R_W=1.23\times A^{1/3}\,\text{fm}$ and adopt the PDG values for $s^2_W$ and $g_A$~\cite{ParticleDataGroup:2024cfk}.
Because of the strong suppression, increasing the fiducial mass has little impact on the relative size of the axial contribution. For example, in tonne-scale detectors the absolute number of axial events becomes sizable, yet it remains overwhelmingly dominated by the vector component. Lowering the recoil-energy threshold does not lead to a significant enhancement of the axial rate either. This overall rescaling is applicable to future multi-ton DM detectors as well, as the one envisaged by the XLZD consortium~\cite{XLZD:2024nsu} that aims at a 60-80 tonne xenon detector. 
\begin{table}[h!]
    \centering
    \setlength{\tabcolsep}{7pt}
    \renewcommand{\arraystretch}{1.5}
    \begin{tabular}{|c||c|c|c|c|}\hline
         Component & CsI & NaI & $^{131}$Xe & $^{73}$Ge\\\hline\hline
         Vector & 39176 & 33944 & 39773 & 25967\\\hline
         Axial & 5 & 9 & 4 & 17\\\hline
    \end{tabular}
    \caption{Expected vector and axial event rates for heavy targets assuming a $50$ kg fiducial mass detector with a $2$ keV recoil threshold and a flat $80\%$ acceptance. Neutrinos are produced via $\pi^+$ decay-at-rest.}
    \label{tab:heavy_compunds_coh_vs_axial.}
    \label{tab:heavy_compunds_coh_vs_axial}
\end{table}

Given these results, target materials suitable for axial-current measurements at \cevns~ facilities should involve relatively light nuclei. A non-exhaustive list includes three $J^\Pi=1/2^+$ isotopes, namely $^3\text{He}$, $^{19}\text{F}$ and $^{29}\text{Si}$, together with deuterium ($J^\Pi=1^+$) and $^{27}\text{Al}$ ($J^\Pi=5/2^+$). Both $^3\text{He}$ and deuterium have a strongly suppressed vector contribution, thus in principle are attractive targets. However, the natural abundance of $^3\text{He}$ is at the parts-per-million level and its low density would require substantially larger fiducial detector masses to achieve meaningful statistics. Deuterium, on the other hand, has already been successfully employed in solar neutrino physics measurements by SNO~\cite{SNO:2005oxr}. 
Currently, the COHERENT collaboration operates a $\sim 500\,\text{kg}$ $\text{D}_2\text{O}$ detector aimed at reducing neutrino-flux statistical uncertainties~\cite{COHERENT:2021xhx}. This detector exploits $\nu_e-\text{D}_2$ interactions observed through Cherenkov light. Although we do not discuss it further, deuterium constitutes a potentially interesting target material for axial-current measurements. Its practical implementation, however, may involve non-trivial technical challenges.

\begin{figure*}[t]
  \centering
    \includegraphics[scale=0.61]{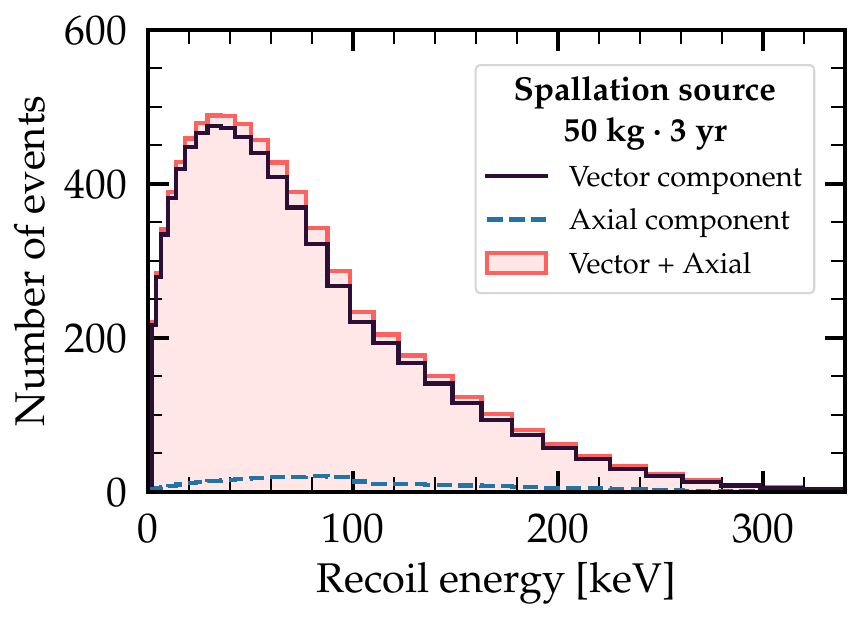}
    \includegraphics[scale=0.61]{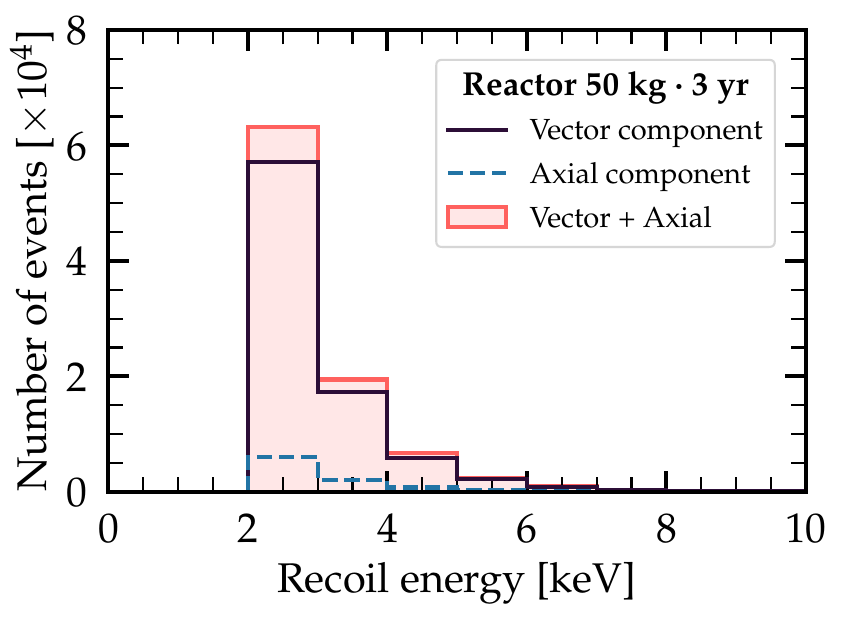}
  \caption{\textbf{Left graph}: Total expected event rate as a function of nuclear recoil energy assuming a spallation-source neutrino flux, for a  $\text{C}_3\text{F}_8$ detector. Following ESS expectations \cite{Baxter:2019mcx}, a recoil-energy threshold of 2 keV is adopted. 
\textbf{Right graph}: Same as the left graph, but for reactor electron antineutrinos. Experimental specifications are taken from the latest CONUS+ run \cite{Ackermann:2025obx} (see text for details).}
  \label{fig:event_rates_C3F8}
\end{figure*}
$^{27}\text{Al}$ appears to be a promising candidate, given its non-zero spin nuclear ground state and its natural abundance. However, we are not aware of detector technologies currently employing this target material. Concerning $^{29}\text{Si}$, the CONNIE experiment~\cite{CONNIE:2019xid} uses silicon detectors, but with strongly limited fiducial masses. Although lower recoil-energy thresholds could, in principle, enhance the axial event rate, the expected signal remains too small to be observable with present configurations. 
As a consequence, among the candidate targets discussed above, fluorine-based materials emerge as the most viable option. Two compounds are particularly relevant: $\text{C}\text{F}_4$ and $\text{C}_3\text{F}_8$. The former has been used in directional DM searches by the DRIFT and the NEWAGE-3.0 collaborations~\cite{Daw:2011wq,Miuchi:2010hn}. It has been proposed as well for \cevns~ measurements at neutrino beam facilities at Fermilab using the $\nu$BDX-DRIFT detector \cite{AristizabalSierra:2021uob,AristizabalSierra:2022jgg}. It has been further considered for DM directional searches by the CYGNO collaboration  \cite{Baracchini:2020btb}.

The $\nu$BDX-DRIFT detector is a gaseous low-pressure TPC where operation thresholds are strongly correlated with operation pressure. In a $1\,\text{m}^3$ detector at $400$ torr, where the \cevns~ signal peaks, the fiducial mass is only $\sim 2$ kg. As a result, minimal detector configurations of this type are unlikely to achieve sufficient sensitivity to axial-current event rates.
Note that, in principle, $\text{C}_4\text{F}_{10}$ could also serve as a target material, following the operating principle of traditional bubble chamber detectors. A notable example is the PICASSO DM experiment~\cite{Archambault:2009sm,PICASSO:2012ngj}, although it operated with a relatively modest fiducial mass. If the same detector configuration could be achievable, in terms of exposure and threshold, we could expect comparable sensitivities to those obtained in the following with $\text{C}_3\text{F}_8$.

The case of $\text{C}_3\text{F}_8$ is indeed qualitatively different, based on experience from the PICO Collaboration~\cite{PICO:2019vsc}.
This makes $\text{C}_3\text{F}_8$ a particularly promising target for axial-current measurements. Motivated by these considerations, we have calculated event rates assuming the same experimental configurations that were employed in the heavy target analysis. In addition, we evaluate the reactor-neutrino case using a configuration resembling that of CONUS+ \cite{Ackermann:2025obx}. The results are presented in Fig.~\ref{fig:event_rates_C3F8}.  Although the axial contribution remains subdominant, its suppression is significantly weaker than in heavy targets. For spallation-source neutrinos we find a suppression factor of $\sim 25$, while for reactor neutrinos it is reduced to $\sim 10$. We attribute this difference to the interplay between the \enquote{morphology} of neutrino spectra and the momentum-transfer dependence of the nuclear responses. Neutrino fluxes at spallation sources peak at energies where the axial spin structure functions decrease more rapidly with momentum transfer than the vector weak-charge form factor. Reactor neutrinos, in contrast, are predominantly produced at lower energies, where the suppression induced by the spin structure functions is milder, resulting in a comparatively enhanced axial contribution.

\textit{Extraction of the axial-vector coupling}---Although the axial-to-vector event ratio remains relatively small, it nevertheless enables studies that would be inaccessible if only the vector component is measured. With data of this type, the most immediate analysis would be the extraction of the axial-vector coupling. In addition, searches for spin-dependent new physics could also be performed. It is worth noting that the absence of observable new physics signals may arise from intrinsically suppressed spin-dependent effects rather than from the absence of new interactions. In the following, we present projected sensitivities to $g_A$ assuming forecasted data. For our analysis, we simulate forecasted data assuming a SM signal. We then allow $g_A$ to vary in the range $[0.8,\,1.6]$, while keeping $g_A^s$ fixed, and construct the following spectral $\chi^2$ function 

\begin{equation}
    \label{eq:chi2_function}
    \chi^2 = 2\sum_i \left[N_i^\text{th} - N_i^\text{exp} + N_i^\text{exp}\ln\left(\frac{N_i^\text{exp}}{N_i^\text{th}}\right)\right] + \left(\frac{\alpha}{\sigma_\alpha}\right)^2\ .
\end{equation}
Here, $N_i^\text{th}$ denotes the theoretically predicted number of events per recoil-energy bin, computed for the reference values $g_A=1.27641$~\cite{Markisch:2018ndu} and $g_A^s=-0.085$ \cite{HERMES:2006jyl}, while $N_i^\text{exp}$ corresponds to the predicted event numbers obtained for varying values of $g_A$. 
The parameter $\alpha$ represents a nuisance parameter accounting for the uncertainty in the neutrino flux, with fractional uncertainty $\sigma_\alpha$.
We consider two representative assumptions for the flux normalization uncertainty: a realistic scenario with $\sigma_\alpha = 5\%$ and an optimistic scenario with $\sigma_\alpha = 1\%$. Although a $1\%$ uncertainty is ambitious, it may not be completely unrealistic\footnote{For example, the COHERENT collaboration expects to determine the pion decay-at-rest neutrino flux with an uncertainty of about $2$-$3\%$ using a dedicated $\mathrm{D_2O}$ detector~\cite{COHERENT:2021xhx}. For the case of reactor antineutrinos, an uncertainty close to 1\%  for the  $^{235}\mathrm{U}$ cross section/fission is reported from the recent Daya Bay's measurement~\cite{DayaBay:2025ngb}, as well as from a global fit~\cite{Giunti:2017nww}.}.
We study two experimental configurations: a near-term detector with a 50 kg fiducial mass operating at recoil-energy thresholds $E_r^\text{th}=3\,$keV and $E_r^\text{th}=2\,$keV, and a future 350 kg phase operating at the same thresholds.  Such exposures appear realistic in light of the planned construction of a $\sim 500$ kg $\text{C}_3\text{F}_8$ detector by the PICO collaboration~\cite{Garcia-Viltres:2021swf}. 

\begin{figure*}[t] %
  \centering
    \includegraphics[scale=0.57]{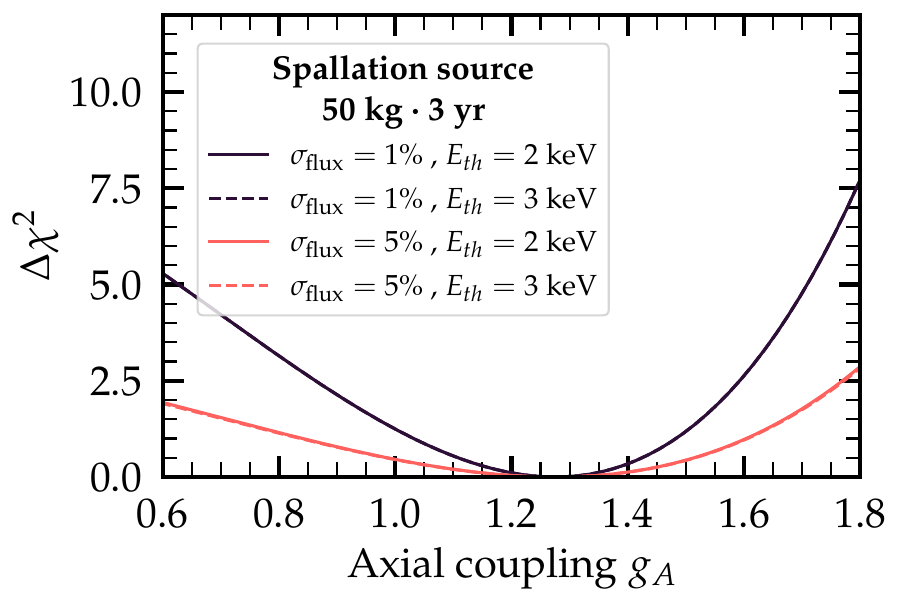}
    \includegraphics[scale=0.57]{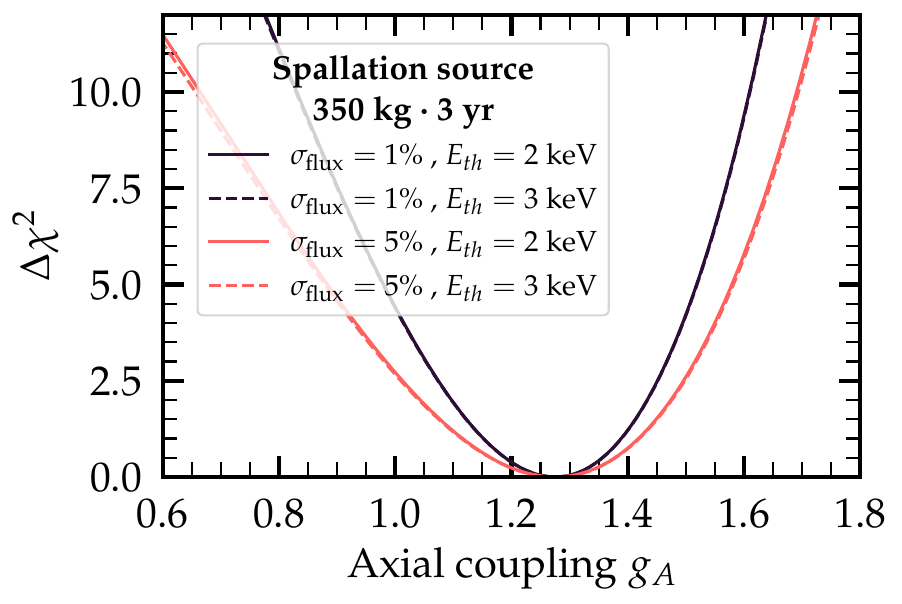}
    \includegraphics[scale=0.57]{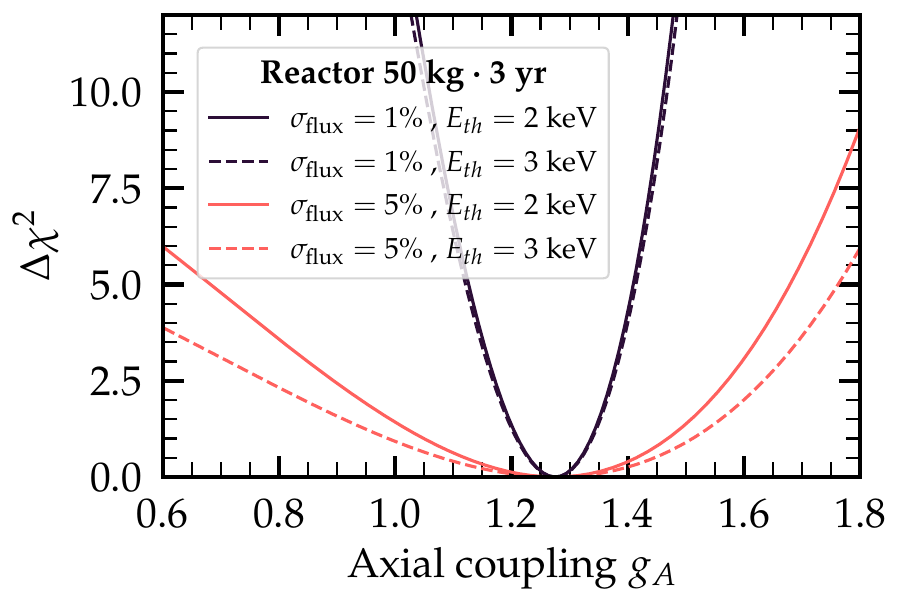}
    \includegraphics[scale=0.57]{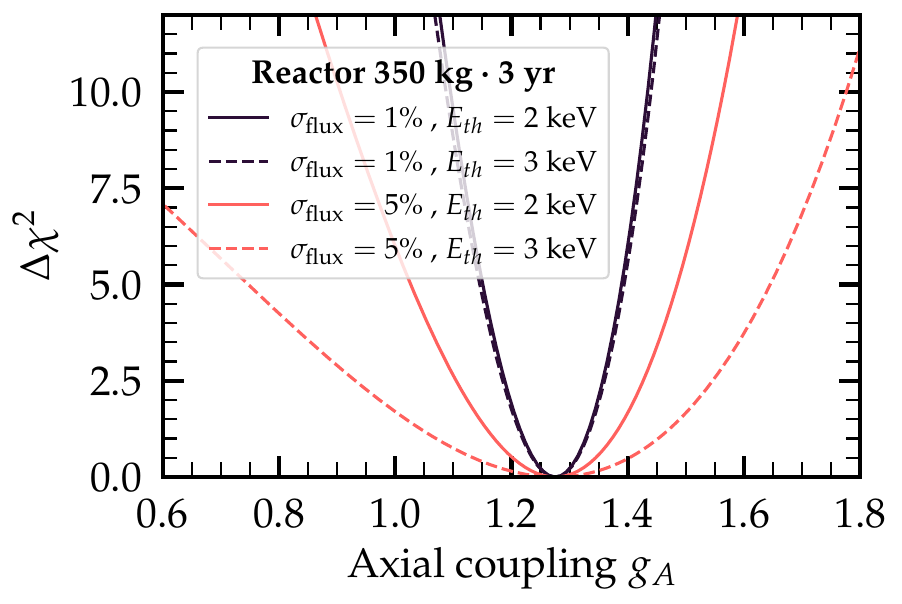}
  \caption{\textbf{Upper panel}: Projected sensitivities to $g_A$, shown as reduced $\chi^2$ profiles, for a spallation-source experiment assuming a $\text{C}_3\text{F}_8$ detector with fiducial masses of 50 kg (\textbf{left}) and 350 kg (\textbf{right}), a 20 m baseline, and different recoil-energy thresholds and neutrino flux uncertainties. \textbf{Lower panel}: Same as the upper panel, but assuming a neutrino flux from a nuclear reactor.}
  \label{fig:chi2_ga}
\end{figure*}

We compute the expected sensitivity for both pion decay-at-rest and reactor neutrino fluxes. The resulting constraints are shown in Fig.~\ref{fig:chi2_ga}.
From the results one can see that when assuming a $3$ keV recoil-energy threshold, neutrino flux uncertainties play a pivotal role. Reducing these uncertainties leads to a substantial improvement in the precision with which $g_A$ can be extracted. In contrast, at a lower threshold ($E_\text{thr} =2$ keV), improvements in the flux uncertainty translate into only marginal gains in precision. Increasing the fiducial mass, however, has a more significant impact, exceeding the improvements obtained solely from reducing neutrino flux uncertainties.
As expected, the optimal determination of $g_A$ is achieved for a combination of lowest threshold, smallest neutrino flux uncertainty, and largest fiducial mass. According to current PICO plans, the benchmark configurations adopted in this work appear realistic and would allow a measurement of $g_A$ at the $\sim 10\%$ level. Achieving a substantially better precision will require simultaneous improvements in several experimental parameters, in particular larger fiducial masses and lower recoil-energy thresholds. Note that attempts in line with this analysis have been done using charged-current NaI COHERENT data in Ref. \cite{Hellgren:2025rap}.

\textit{Conclusions}---The axial neutral-current provides a subdominant contribution to \cevns~event rates in targets with a non-zero nuclear ground-state spin. In counting experiments, its signal is largely overwhelmed by the dominant vector component, which explains why it has so far eluded experimental observation, and justifies its neglect in phenomenological analyses of current measurements, obtained with heavy-target materials. Nevertheless, accessing this contribution is essential for fully exploiting the physics potential of the \cevns~physics agenda.

In this letter, we have investigated the prospects for measuring this subleading yet phenomenologically important component. We have first examined heavy target materials commonly employed in neutrino and DM experiments and demonstrated that, even with tonne-scale detectors, the axial-current contribution remains effectively negligible. Although a small number of axial events may be collected, the statistical information that can be extracted is limited. We then have identified medium-mass and light nuclei with non-zero spin ground states as the more promising candidates. For several of these isotopes, however, suitable detector technologies are either unavailable or not currently planned. 

These considerations lead us to conclude that fluorine-based compounds provide the most suitable targets, with $\text{C}\text{F}_4$ and $\text{C}_3\text{F}_8$ emerging as the most promising options. Among them, $\text{C}_3\text{F}_8$ stands out due to its extensive use over nearly two decades in DM direct detection searches and the proposed deployment of detectors based on this material at the ESS~\cite{Baxter:2019mcx} and at J-PARC~\cite{Collar:2025sle} facilities.

As an application, we performed a forecast study of axial-current measurements and evaluated the achievable precision on the extraction of the axial-vector coupling $g_A$. We find that in a realistic experimental setup consisting of 350 kg of $\text{C}_3\text{F}_8$, a 2~keV nuclear-recoil threshold and a $1\%$ neutrino flux uncertainty, an extraction at the $\sim 10\%$ precision level seems possible. Achieving significantly better precision, would require simultaneous improvements in detector threshold, fiducial mass and neutrino flux determination, which may prove experimentally challenging.
These measurements will be complementary to those done in dedicated experiments, thus providing further insight on $g_A$. Finally, we emphasize that measurements of the axial-current contribution would open a new avenue for the study of low-energy new-physics signals that may remain hidden in standard analysis due to the intrinsic suppression of spin-dependent effects.

\textit{Acknowledgments}---P.M.C and V.D.R. acknowledge financial support by the grant CIDEXG/2022/20 (from Generalitat Valenciana) and by the Spanish grants CNS2023-144124 (MCIN/AEI/10.13039/501100011033 and “Next Generation EU”/PRTR), PID2023-147306NB-I00, and CEX2023-001292-S (MCIU/AEI/10.13039/ 501100011033). P.M.C. is also supported by the grant CIACIF/2021/281 (Generalitat Valenciana). The work of D.A.S. and L.T.S is funded by
ANID grants \enquote{Fondecyt Regular} 1221445 and 1260595. L.T.S is funded as well by ``Programa de Iniciaci\'on a la Investigaci\'on Cient\'{\i}fica (PIIC)'' number 007/2025. Both thank IFIC and specially the members of the AHEP group for their kind hospitality during several stages of this project.  D.K.P. acknowledges funding from the European Union’s Horizon Europe research and innovation programme under the Marie Skłodowska‑Curie Actions grant agreement No.~101198541 (neutrinoSPHERE).

\bibliography{bibliography}

@article{PandaX:2024muv,
    author = "Bo, Zihao and others",
    collaboration = "PandaX",
    title = "{First Indication of Solar B8 Neutrinos through Coherent Elastic Neutrino-Nucleus Scattering in PandaX-4T}",
    eprint = "2407.10892",
    archivePrefix = "arXiv",
    primaryClass = "hep-ex",
    doi = "10.1103/PhysRevLett.133.191001",
    journal = "Phys. Rev. Lett.",
    volume = "133",
    number = "19",
    pages = "191001",
    year = "2024"
}

@article{HERMES:2006jyl,
    author = "Airapetian, A. and others",
    collaboration = "HERMES",
    title = "{Precise determination of the spin structure function g(1) of the proton, deuteron and neutron}",
    eprint = "hep-ex/0609039",
    archivePrefix = "arXiv",
    reportNumber = "DESY-06-142",
    doi = "10.1103/PhysRevD.75.012007",
    journal = "Phys. Rev. D",
    volume = "75",
    pages = "012007",
    year = "2007"
}

@article{AristizabalSierra:2021uob,
    author = "Aristizabal Sierra, D. and Dutta, Bhaskar and Kim, Doojin and Snowden-Ifft, Daniel and Strigari, Louis E.",
    title = "{Coherent elastic neutrino-nucleus scattering with the {\ensuremath{\nu}}BDX-DRIFT directional detector at next generation neutrino facilities}",
    eprint = "2103.10857",
    archivePrefix = "arXiv",
    primaryClass = "hep-ph",
    doi = "10.1103/PhysRevD.104.033004",
    journal = "Phys. Rev. D",
    volume = "104",
    number = "3",
    pages = "033004",
    year = "2021"
}

@article{AristizabalSierra:2022jgg,
    author = "Aristizabal Sierra, D. and Barrow, J. L. and Dutta, B. and Kim, D. and Snowden-Ifft, D. and Strigari, L. and Wood, M. H.",
    collaboration = "{\ensuremath{\nu}}BDX-DRIFT",
    title = "{Rock neutron backgrounds from FNAL neutrino beamlines in the {\ensuremath{\nu}}BDX-DRIFT detector}",
    eprint = "2210.08612",
    archivePrefix = "arXiv",
    primaryClass = "hep-ex",
    doi = "10.1103/PhysRevD.107.013003",
    journal = "Phys. Rev. D",
    volume = "107",
    number = "1",
    pages = "013003",
    year = "2023"
}

@article{SNO:2005oxr,
    author = "Aharmim, B. and others",
    collaboration = "SNO",
    title = "{Electron energy spectra, fluxes, and day-night asymmetries of B-8 solar neutrinos from measurements with NaCl dissolved in the heavy-water detector at the Sudbury Neutrino Observatory}",
    eprint = "nucl-ex/0502021",
    archivePrefix = "arXiv",
    doi = "10.1103/PhysRevC.72.055502",
    journal = "Phys. Rev. C",
    volume = "72",
    pages = "055502",
    year = "2005"
}

@article{Majorana:2013cem,
    author = "Abgrall, N. and others",
    collaboration = "Majorana",
    title = "{The Majorana Demonstrator Neutrinoless Double-Beta Decay Experiment}",
    eprint = "1308.1633",
    archivePrefix = "arXiv",
    primaryClass = "physics.ins-det",
    reportNumber = "LA-UR-13-24815",
    doi = "10.1155/2014/365432",
    journal = "Adv. High Energy Phys.",
    volume = "2014",
    pages = "365432",
    year = "2014"
}

@article{GERDA:2020xhi,
    author = "Agostini, M. and others",
    collaboration = "GERDA",
    title = "{Final Results of GERDA on the Search for Neutrinoless Double-$\beta$ Decay}",
    eprint = "2009.06079",
    archivePrefix = "arXiv",
    primaryClass = "nucl-ex",
    doi = "10.1103/PhysRevLett.125.252502",
    journal = "Phys. Rev. Lett.",
    volume = "125",
    number = "25",
    pages = "252502",
    year = "2020"
}

@article{Papoulias:2019lfi,
    author = "Papoulias, D. K. and Kosmas, T. S. and Sahu, R. and Kota, V. K. B. and Hota, M.",
    title = "{Constraining nuclear physics parameters with current and future COHERENT data}",
    eprint = "1903.03722",
    archivePrefix = "arXiv",
    primaryClass = "hep-ph",
    reportNumber = "IFIC/19-xxx",
    doi = "10.1016/j.physletb.2019.135133",
    journal = "Phys. Lett. B",
    volume = "800",
    pages = "135133",
    year = "2020"
}

@article{DayaBay:2025ngb,
    author = "An, F. P. and others",
    collaboration = "Daya Bay",
    title = "{Comprehensive Measurement of the Reactor Antineutrino Spectrum and Flux at Daya Bay}",
    eprint = "2501.00746",
    archivePrefix = "arXiv",
    primaryClass = "nucl-ex",
    doi = "10.1103/PhysRevLett.134.201802",
    journal = "Phys. Rev. Lett.",
    volume = "134",
    number = "20",
    pages = "201802",
    year = "2025"
}

@article{Giunti:2017nww,
    author = "Giunti, Carlo",
    title = "{Improved Determination of the $^{235}\text{U}$ and $^{239}\text{Pu}$ Reactor Antineutrino Cross Sections per Fission}",
    eprint = "1704.02276",
    archivePrefix = "arXiv",
    primaryClass = "hep-ph",
    doi = "10.1103/PhysRevD.96.033005",
    journal = "Phys. Rev. D",
    volume = "96",
    number = "3",
    pages = "033005",
    year = "2017"
}

@article{DelNobile:2021wmp,
    author = "Del Nobile, Eugenio",
    title = "{The Theory of Direct Dark Matter Detection: A Guide to Computations}",
    eprint = "2104.12785",
    archivePrefix = "arXiv",
    primaryClass = "hep-ph",
    doi = "10.1007/978-3-030-95228-0",
    month = "4",
    year = "2021"
}

@article{Belanger:2008sj,
    author = "Belanger, G. and Boudjema, F. and Pukhov, A. and Semenov, A.",
    title = "{Dark matter direct detection rate in a generic model with micrOMEGAs 2.2}",
    eprint = "0803.2360",
    archivePrefix = "arXiv",
    primaryClass = "hep-ph",
    reportNumber = "LAPTH-1237-08",
    doi = "10.1016/j.cpc.2008.11.019",
    journal = "Comput. Phys. Commun.",
    volume = "180",
    pages = "747--767",
    year = "2009"
}

@article{Cerdeno:2012ix,
    author = "Cerdeno, D. G. and Fornasa, M. and Huh, J. -H. and Peiro, M.",
    title = "{Nuclear uncertainties in the spin-dependent structure functions for direct dark matter detection}",
    eprint = "1208.6426",
    archivePrefix = "arXiv",
    primaryClass = "hep-ph",
    reportNumber = "FTUAM-12-101, IFT-UAM-CSIC-12-86",
    doi = "10.1103/PhysRevD.87.023512",
    journal = "Phys. Rev. D",
    volume = "87",
    number = "2",
    pages = "023512",
    year = "2013"
}

@article{COHERENT:2021xhx,
    author = "Akimov, D. and others",
    collaboration = "COHERENT",
    title = "{A D$_2$O detector for flux normalization of a pion decay-at-rest neutrino source}",
    eprint = "2104.09605",
    archivePrefix = "arXiv",
    primaryClass = "physics.ins-det",
    reportNumber = "FERMILAB-PUB-21-208-ND",
    doi = "10.1088/1748-0221/16/08/P08048",
    journal = "JINST",
    volume = "16",
    number = "08",
    pages = "P08048",
    year = "2021"
}

@article{CONNIE:2019xid,
    author = "Aguilar-Arevalo, Alexis and others",
    collaboration = "CONNIE",
    title = "{Search for light mediators in the low-energy data of the CONNIE reactor neutrino experiment}",
    eprint = "1910.04951",
    archivePrefix = "arXiv",
    primaryClass = "hep-ex",
    reportNumber = "FERMILAB-PUB-19-525-AE-PPD-SCD",
    doi = "10.1007/JHEP04(2020)054",
    journal = "JHEP",
    volume = "04",
    pages = "054",
    year = "2020"
}

@article{Baxter:2019mcx,
    author = "Baxter, D. and others",
    title = "{Coherent Elastic Neutrino-Nucleus Scattering at the European Spallation Source}",
    eprint = "1911.00762",
    archivePrefix = "arXiv",
    primaryClass = "physics.ins-det",
    reportNumber = "IFIC/19-45, YITP-SB-19-37, FERMILAB-PUB-19-612-V",
    doi = "10.1007/JHEP02(2020)123",
    journal = "JHEP",
    volume = "02",
    pages = "123",
    year = "2020"
}

@article{Garcia-Viltres:2021swf,
    author = "Garc{\'\i}a-Viltres, A. S. and V{\'a}zquez-J{\'a}uregui, E.",
    collaboration = "PICO",
    title = "{PICO-500: A tonne scale bubble chamber for the search of dark matter}",
    doi = "10.1393/ncc/i2022-22007-x",
    journal = "Nuovo Cim. C",
    volume = "45",
    number = "1",
    pages = "7",
    year = "2021"
}

@article{PICO:2017tgi,
    author = "Amole, C. and others",
    collaboration = "PICO",
    title = "{Dark Matter Search Results from the PICO-60 C$_3$F$_8$ Bubble Chamber}",
    eprint = "1702.07666",
    archivePrefix = "arXiv",
    primaryClass = "astro-ph.CO",
    reportNumber = "FERMILAB-PUB-17-058-AE-PPD",
    doi = "10.1103/PhysRevLett.118.251301",
    journal = "Phys. Rev. Lett.",
    volume = "118",
    number = "25",
    pages = "251301",
    year = "2017"
}

@article{LZ:2025igz,
    author = "Akerib, D. S. and others",
    collaboration = "LZ",
    title = "{Searches for Light Dark Matter and Evidence of Coherent Elastic Neutrino-Nucleus Scattering of Solar Neutrinos with the LUX-ZEPLIN (LZ) Experiment}",
    eprint = "2512.08065",
    archivePrefix = "arXiv",
    primaryClass = "hep-ex",
    month = "12",
    year = "2025"
}

@article{Hoferichter:2020osn,
    author = "Hoferichter, Martin and Men{\'e}ndez, Javier and Schwenk, Achim",
    title = "{Coherent elastic neutrino-nucleus scattering: EFT analysis and nuclear responses}",
    eprint = "2007.08529",
    archivePrefix = "arXiv",
    primaryClass = "hep-ph",
    reportNumber = "INT-PUB-20-026",
    doi = "10.1103/PhysRevD.102.074018",
    journal = "Phys. Rev. D",
    volume = "102",
    number = "7",
    pages = "074018",
    year = "2020"
}

@article{Klos:2013rwa,
    author = "Klos, P. and Men{\'e}ndez, J. and Gazit, D. and Schwenk, A.",
    title = "{Large-scale nuclear structure calculations for spin-dependent WIMP scattering with chiral effective field theory currents}",
    eprint = "1304.7684",
    archivePrefix = "arXiv",
    primaryClass = "nucl-th",
    doi = "10.1103/PhysRevD.88.083516",
    journal = "Phys. Rev. D",
    volume = "88",
    number = "8",
    pages = "083516",
    year = "2013",
    note = "[Erratum: Phys.Rev.D 89, 029901 (2014)]"
}

@article{Markisch:2018ndu,
    author = {M{\"a}rkisch, B. and others},
    title = "{Measurement of the Weak Axial-Vector Coupling Constant in the Decay of Free Neutrons Using a Pulsed Cold Neutron Beam}",
    eprint = "1812.04666",
    archivePrefix = "arXiv",
    primaryClass = "nucl-ex",
    doi = "10.1103/PhysRevLett.122.242501",
    journal = "Phys. Rev. Lett.",
    volume = "122",
    number = "24",
    pages = "242501",
    year = "2019"
}

@article{ParticleDataGroup:2024cfk,
    author = "Navas, S. and others",
    collaboration = "Particle Data Group",
    title = "{Review of particle physics}",
    doi = "10.1103/PhysRevD.110.030001",
    journal = "Phys. Rev. D",
    volume = "110",
    number = "3",
    pages = "030001",
    year = "2024"
}

@article{COHERENT:2017ipa,
    author = "Akimov, D. and others",
    collaboration = "COHERENT",
    title = "{Observation of Coherent Elastic Neutrino-Nucleus Scattering}",
    eprint = "1708.01294",
    archivePrefix = "arXiv",
    primaryClass = "nucl-ex",
    doi = "10.1126/science.aao0990",
    journal = "Science",
    volume = "357",
    number = "6356",
    pages = "1123--1126",
    year = "2017"
}

@article{DeRomeri:2022twg,
    author = "De Romeri, V. and Miranda, O. G. and Papoulias, D. K. and Sanchez Garcia, G. and T{\'o}rtola, M. and Valle, J. W. F.",
    title = "{Physics implications of a combined analysis of COHERENT CsI and LAr data}",
    eprint = "2211.11905",
    archivePrefix = "arXiv",
    primaryClass = "hep-ph",
    doi = "10.1007/JHEP04(2023)035",
    journal = "JHEP",
    volume = "04",
    pages = "035",
    year = "2023"
}

@article{Cadeddu:2020lky,
    author = "Cadeddu, M. and Dordei, F. and Giunti, C. and Li, Y. F. and Picciau, E. and Zhang, Y. Y.",
    title = "{Physics results from the first COHERENT observation of coherent elastic neutrino-nucleus scattering in argon and their combination with cesium-iodide data}",
    eprint = "2005.01645",
    archivePrefix = "arXiv",
    primaryClass = "hep-ph",
    doi = "10.1103/PhysRevD.102.015030",
    journal = "Phys. Rev. D",
    volume = "102",
    number = "1",
    pages = "015030",
    year = "2020"
}

@article{COHERENT:2020iec,
    author = "Akimov, D. and others",
    collaboration = "COHERENT",
    title = "{First Measurement of Coherent Elastic Neutrino-Nucleus Scattering on Argon}",
    eprint = "2003.10630",
    archivePrefix = "arXiv",
    primaryClass = "nucl-ex",
    doi = "10.1103/PhysRevLett.126.012002",
    journal = "Phys. Rev. Lett.",
    volume = "126",
    number = "1",
    pages = "012002",
    year = "2021"
}

@article{COHERENT:2021xmm,
    author = "Akimov, D. and others",
    collaboration = "COHERENT",
    title = "{Measurement of the Coherent Elastic Neutrino-Nucleus Scattering Cross Section on CsI by COHERENT}",
    eprint = "2110.07730",
    archivePrefix = "arXiv",
    primaryClass = "hep-ex",
    doi = "10.1103/PhysRevLett.129.081801",
    journal = "Phys. Rev. Lett.",
    volume = "129",
    number = "8",
    pages = "081801",
    year = "2022"
}

@article{COHERENT:2024axu,
    author = "Adamski, S. and others",
    collaboration = "COHERENT",
    title = "{First detection of coherent elastic neutrino-nucleus scattering on germanium}",
    eprint = "2406.13806",
    archivePrefix = "arXiv",
    primaryClass = "hep-ex",
    month = "6",
    year = "2024"
}

@article{Ackermann:2025obx,
    author = "Ackermann, N. and others",
    title = "{Direct observation of coherent elastic antineutrino{\textendash}nucleus scattering}",
    eprint = "2501.05206",
    archivePrefix = "arXiv",
    primaryClass = "hep-ex",
    doi = "10.1038/s41586-025-09322-2",
    journal = "Nature",
    volume = "643",
    number = "8074",
    pages = "1229--1233",
    year = "2025"
}

@article{Barranco:2005yy,
    author = "Barranco, J. and Miranda, O. G. and Rashba, T. I.",
    title = "{Probing new physics with coherent neutrino scattering off nuclei}",
    eprint = "hep-ph/0508299",
    archivePrefix = "arXiv",
    reportNumber = "MPP-2005-85",
    doi = "10.1088/1126-6708/2005/12/021",
    journal = "JHEP",
    volume = "12",
    pages = "021",
    year = "2005"
}

@article{Helm:1956zz,
    author = "Helm, Richard H.",
    title = "{Inelastic and Elastic Scattering of 187-Mev Electrons from Selected Even-Even Nuclei}",
    doi = "10.1103/PhysRev.104.1466",
    journal = "Phys. Rev.",
    volume = "104",
    pages = "1466--1475",
    year = "1956"
}

@article{Klein:1999qj,
    author = "Klein, Spencer and Nystrand, Joakim",
    title = "{Exclusive vector meson production in relativistic heavy ion collisions}",
    eprint = "hep-ph/9902259",
    archivePrefix = "arXiv",
    reportNumber = "LBNL-42768, LBL-42768",
    doi = "10.1103/PhysRevC.60.014903",
    journal = "Phys. Rev. C",
    volume = "60",
    pages = "014903",
    year = "1999"
}

@article{XENON:2024ijk,
    author = "Aprile, Elena and others",
    collaboration = "XENON",
    title = "{First Indication of Solar B8 Neutrinos via Coherent Elastic Neutrino-Nucleus Scattering with XENONnT}",
    eprint = "2408.02877",
    archivePrefix = "arXiv",
    primaryClass = "nucl-ex",
    doi = "10.1103/PhysRevLett.133.191002",
    journal = "Phys. Rev. Lett.",
    volume = "133",
    number = "19",
    pages = "191002",
    year = "2024"
}

@article{Colaresi:2022obx,
    author = "Colaresi, J. and Collar, J. I. and Hossbach, T. W. and Lewis, C. M. and Yocum, K. M.",
    title = "{Measurement of Coherent Elastic Neutrino-Nucleus Scattering from Reactor Antineutrinos}",
    eprint = "2202.09672",
    archivePrefix = "arXiv",
    primaryClass = "hep-ex",
    doi = "10.1103/PhysRevLett.129.211802",
    journal = "Phys. Rev. Lett.",
    volume = "129",
    number = "21",
    pages = "211802",
    year = "2022"
}

@article{Freedman:1973yd,
    author = "Freedman, Daniel Z.",
    title = "{Coherent Neutrino Nucleus Scattering as a Probe of the Weak Neutral Current}",
    reportNumber = "NAL-PUB-73-76-THY, FERMILAB-PUB-73-076-T",
    doi = "10.1103/PhysRevD.9.1389",
    journal = "Phys. Rev. D",
    volume = "9",
    pages = "1389--1392",
    year = "1974"
}

@article{Abdullah:2022zue,
    author = "Abdullah, M. and others",
    title = "{Coherent elastic neutrino-nucleus scattering: Terrestrial and astrophysical applications}",
    eprint = "2203.07361",
    archivePrefix = "arXiv",
    primaryClass = "hep-ph",
    month = "3",
    year = "2022"
}

@article{Schumann:2019eaa,
    author = "Schumann, Marc",
    title = "{Direct Detection of WIMP Dark Matter: Concepts and Status}",
    eprint = "1903.03026",
    archivePrefix = "arXiv",
    primaryClass = "astro-ph.CO",
    doi = "10.1088/1361-6471/ab2ea5",
    journal = "J. Phys. G",
    volume = "46",
    number = "10",
    pages = "103003",
    year = "2019"
}

@article{MarrodanUndagoitia:2015veg,
    author = "Marrod{\'a}n Undagoitia, Teresa and Rauch, Ludwig",
    title = "{Dark matter direct-detection experiments}",
    eprint = "1509.08767",
    archivePrefix = "arXiv",
    primaryClass = "physics.ins-det",
    doi = "10.1088/0954-3899/43/1/013001",
    journal = "J. Phys. G",
    volume = "43",
    number = "1",
    pages = "013001",
    year = "2016"
}

@article{Billard:2021uyg,
    author = "Billard, Julien and others",
    title = "{Direct detection of dark matter{\textemdash}APPEC committee report*}",
    eprint = "2104.07634",
    archivePrefix = "arXiv",
    primaryClass = "hep-ex",
    doi = "10.1088/1361-6633/ac5754",
    journal = "Rept. Prog. Phys.",
    volume = "85",
    number = "5",
    pages = "056201",
    year = "2022"
}

@article{PICO:2019vsc,
    author = "Amole, C. and others",
    collaboration = "PICO",
    title = "{Dark Matter Search Results from the Complete Exposure of the PICO-60 C$_3$F$_8$ Bubble Chamber}",
    eprint = "1902.04031",
    archivePrefix = "arXiv",
    primaryClass = "astro-ph.CO",
    reportNumber = "FERMILAB-PUB-19-073-AE-E",
    doi = "10.1103/PhysRevD.100.022001",
    journal = "Phys. Rev. D",
    volume = "100",
    number = "2",
    pages = "022001",
    year = "2019"
}

@article{PICO:2019rsv,
    author = "Amole, C. and others",
    collaboration = "PICO",
    title = "{Data-Driven Modeling of Electron Recoil Nucleation in PICO C$_3$F$_8$ Bubble Chambers}",
    eprint = "1905.12522",
    archivePrefix = "arXiv",
    primaryClass = "physics.ins-det",
    reportNumber = "FERMILAB-PUB-19-254-AE",
    doi = "10.1103/PhysRevD.100.082006",
    journal = "Phys. Rev. D",
    volume = "100",
    number = "8",
    pages = "082006",
    year = "2019"
}

@article{Baxter:2017ozv,
    author = "Baxter, D. and others",
    title = "{First Demonstration of a Scintillating Xenon Bubble Chamber for Detecting Dark Matter and Coherent Elastic Neutrino-Nucleus Scattering}",
    eprint = "1702.08861",
    archivePrefix = "arXiv",
    primaryClass = "physics.ins-det",
    reportNumber = "FERMILAB-PUB-17-062-AE-E-PPD",
    doi = "10.1103/PhysRevLett.118.231301",
    journal = "Phys. Rev. Lett.",
    volume = "118",
    number = "23",
    pages = "231301",
    year = "2017"
}

@article{Collar:2025sle,
    author = "Collar, J. I. and others",
    title = "{Coherent Elastic Neutrino-Nucleus Scattering at the Japan Proton Accelerator Research Complex}",
    eprint = "2512.19788",
    archivePrefix = "arXiv",
    primaryClass = "hep-ph",
    month = "12",
    year = "2025"
}

@article{AtzoriCorona:2024rtv,
    author = "Atzori Corona, M. and Cadeddu, M. and Cargioli, N. and Dordei, F. and Giunti, C.",
    title = "{Momentum dependent flavor radiative corrections to the coherent elastic neutrino-nucleus scattering for the neutrino charge-radius determination}",
    eprint = "2402.16709",
    archivePrefix = "arXiv",
    primaryClass = "hep-ph",
    doi = "10.1007/JHEP05(2024)271",
    journal = "JHEP",
    volume = "05",
    pages = "271",
    year = "2024"
}

@article{Mishra:2023jlq,
    author = "Mishra, Nityasa and Strigari, Louis E.",
    title = "{Solar neutrinos with CE{\ensuremath{\nu}}NS and flavor-dependent radiative corrections}",
    eprint = "2305.17827",
    archivePrefix = "arXiv",
    primaryClass = "hep-ph",
    reportNumber = "MI-HET-812",
    doi = "10.1103/PhysRevD.108.063023",
    journal = "Phys. Rev. D",
    volume = "108",
    number = "6",
    pages = "063023",
    year = "2023"
}

@article{Tomalak:2020zfh,
    author = "Tomalak, Oleksandr and Machado, Pedro and Pandey, Vishvas and Plestid, Ryan",
    title = "{Flavor-dependent radiative corrections in coherent elastic neutrino-nucleus scattering}",
    eprint = "2011.05960",
    archivePrefix = "arXiv",
    primaryClass = "hep-ph",
    reportNumber = "FERMILAB-PUB-20-524-T",
    doi = "10.1007/JHEP02(2021)097",
    journal = "JHEP",
    volume = "02",
    pages = "097",
    year = "2021"
}

@article{COHERENT:2023ffx,
    author = "An, P. and others",
    collaboration = "COHERENT",
    title = "{Measurement of Electron-Neutrino Charged-Current Cross Sections on I127 with the COHERENT NaI{\ensuremath{\nu}}E Detector}",
    eprint = "2305.19594",
    archivePrefix = "arXiv",
    primaryClass = "nucl-ex",
    doi = "10.1103/PhysRevLett.131.221801",
    journal = "Phys. Rev. Lett.",
    volume = "131",
    number = "22",
    pages = "221801",
    year = "2023"
}

@article{CSNS,
  author={S. W. Xu and Others},
  title={China Spallation Neutron Source - an overview of application and development},
  journal={Chinese Physics C},
  volume={33},
  number={11},
  pages={1--5},
  year={2009},
  publisher={IOP Publishing},
  doi={10.1088/1674-1137/33/11/021},
url = {https://doi.org/10.1088/1674-1137/33/11/021}}

@article{Su:2023klh,
    author = "Su, Chenguang and Liu, Qian and Liang, Tianjiao",
    collaboration = "CLOVERS, CE{\ensuremath{\nu}}NS@CSNS",
    title = "{CE{\ensuremath{\nu}}NS Experiment Proposal at CSNS {\textdagger}}",
    eprint = "2303.13423",
    archivePrefix = "arXiv",
    primaryClass = "physics.ins-det",
    doi = "10.3390/psf2023008019",
    journal = "Phys. Sci. Forum",
    volume = "8",
    number = "1",
    pages = "19",
    year = "2023"
}

@article{XLZD:2024nsu,
    author = "Aalbers, J. and others",
    collaboration = "XLZD",
    title = "{The XLZD Design Book: towards the next-generation liquid xenon observatory for dark matter and neutrino physics}",
    eprint = "2410.17137",
    archivePrefix = "arXiv",
    primaryClass = "hep-ex",
    doi = "10.1140/epjc/s10052-025-14810-w",
    journal = "Eur. Phys. J. C",
    volume = "85",
    number = "10",
    pages = "1192",
    year = "2025"
}

@article{Archambault:2009sm,
    author = "Archambault, S. and others",
    title = "{Dark Matter Spin-Dependent Limits for WIMP Interactions on F-19 by PICASSO}",
    eprint = "0907.0307",
    archivePrefix = "arXiv",
    primaryClass = "hep-ex",
    doi = "10.1016/j.physletb.2009.11.019",
    journal = "Phys. Lett. B",
    volume = "682",
    pages = "185--192",
    year = "2009"
}

@article{PICASSO:2012ngj,
    author = "Archambault, S. and others",
    collaboration = "PICASSO",
    title = "{Constraints on Low-Mass WIMP Interactions on $^{19}F$ from PICASSO}",
    eprint = "1202.1240",
    archivePrefix = "arXiv",
    primaryClass = "hep-ex",
    doi = "10.1016/j.physletb.2012.03.078",
    journal = "Phys. Lett. B",
    volume = "711",
    pages = "153--161",
    year = "2012"
}

@article{Miuchi:2010hn,
    author = "Miuchi, Kentaro and others",
    title = "{First underground results with NEWAGE-0.3a direction-sensitive dark matter detector}",
    eprint = "1002.1794",
    archivePrefix = "arXiv",
    primaryClass = "astro-ph.CO",
    doi = "10.1016/j.physletb.2010.02.028",
    journal = "Phys. Lett. B",
    volume = "686",
    pages = "11--17",
    year = "2010"
}

@article{Daw:2011wq,
    author = "Daw, E. and others",
    editor = "Mayet, F. and Santos, D.",
    title = "{The DRIFT Directional Dark Matter Experiments}",
    eprint = "1110.0222",
    archivePrefix = "arXiv",
    primaryClass = "physics.ins-det",
    doi = "10.1051/eas/1253002",
    journal = "EAS Publ. Ser.",
    volume = "53",
    pages = "11--18",
    year = "2012"
}

@article{Baracchini:2020btb,
    author = "Baracchini, E. and others",
    title = "{CYGNO: a gaseous TPC with optical readout for dark matter directional search}",
    eprint = "2007.12627",
    archivePrefix = "arXiv",
    primaryClass = "physics.ins-det",
    doi = "10.1088/1748-0221/15/07/C07036",
    journal = "JINST",
    volume = "15",
    number = "07",
    pages = "C07036",
    year = "2020"
}

@article{Hellgren:2025rap,
    author = "Hellgren, M. and Suhonen, J.",
    title = "{Quenching of the weak axial coupling derived from the scattering of stopped-pion neutrinos on 127I}",
    doi = "10.1016/j.physletb.2025.139508",
    journal = "Phys. Lett. B",
    volume = "866",
    pages = "139508",
    year = "2025"
}

\begin{widetext}
\appendix

\begin{center}
 \bf \large Supplemental Material
\end{center}
\vspace*{0.2cm}

Neutrinos produced at spallation neutron sources originate primarily from $\pi^+$ meson decays. At the energies relevant for these facilities most $\pi^-$ mesons are efficiently absorbed by nuclei before decaying. As a result, three neutrino flavors are then produced: a prompt $\nu_\mu$ component and delayed $\nu_e$ and $\overline{\nu}_\mu$. The prompt signal, arises from the two-body decay $\pi^+ \to \mu^+ \nu_\mu$ and hence is monochromatic, with energy located at $E_{\nu_\mu}=(m_{\pi^+}^2-m_\mu^2)/2 m_{\pi^+}$. The delayed signal, stemming from a three-body decay mode, involves two continuous spectra dictated by the muon energy spectra. The spectral functions follow from phase space integration
\begin{align}
    \label{eq:spectra_pi0_decay_at_rest}
    \frac{\d\Phi_{\nu_\mu}}{\d E_{\nu_\mu}}&=
    \frac{2m_{\pi^+}}{m_{\pi^+}^2 - m_\mu^2}
    \delta\!\left(
    1 - \frac{2 E_{\nu_\mu} m_{\pi^+}}{m_{\pi^+}^2 - m_\mu^2}
    \right)\ ,
    \nonumber\\
    \frac{\d\Phi_{\overline{\nu}_\mu}}{\d E_{\overline{\nu}_\mu}}&=\frac{64}{m_\mu}
    \left(\frac{E_{\overline{\nu}_\mu}}{m_\mu}\right)^2
    \left(
    \frac{3}{4}-\frac{E_{\overline{\nu}_\mu}}{m_\mu}
    \right) \, \Theta\!\left( E_{\bar{\nu}_\mu}-\frac{m_\mu}{2}\right)\ ,
    \nonumber\\
     \frac{\d\Phi_{\nu_e}}{\d E_{\nu_e}}&=\frac{192}{m_\mu}
    \left(\frac{E_{\nu_e}}{m_\mu}\right)^2
    \left(
    \frac{1}{2}-\frac{E_{\nu_e}}{m_\mu}
    \right) \, \Theta\!\left( E_{\nu_e}-\frac{m_\mu}{2}\right)\ .
\end{align}
Flux normalization is determined by the value of protons on target (POT), baseline and the fraction of neutrinos produced per POT. The latter depends on the target material hit by the protons as well as the proton kinetic energy. We implement results for the ESS, for which plans to deploy a $\text{C}_3\text{F}_8$ detector have been discussed in Ref. \cite{Baxter:2019mcx}. In that case, 0.3 neutrinos of each flavor per POT, assuming a 2 GeV proton-collision energy, are expected~\cite{Baxter:2019mcx}.

The reactor electron anti-neutrino flux is instead implemented according to Ref.~\cite{CONNIE:2019xid}. More than $80\%$ of the flux is generated from $\beta$ decay processes of the fissile isotopes: $^{235}$U, $^{235}$U, $^{239}$Pu and $^{241}$Pu. Above $2\,$MeV, the spectrum is well described by
\begin{equation}
    \label{eq:nu_flux_above_2MeV}
    \frac{\d\Phi_{\overline{\nu}_e}}{\d E_{\overline{\nu}_e}}
    = \frac{\mathcal{P}}{4\pi\, \varepsilon L^2}\sum_i f_i\,
    \frac{\d N^i_{\overline{\nu}_e}}{\d E_{\overline{\nu}_e}}\ ,
\end{equation}
with the fission fractions $f_i$ given by $\{f_{^{235}\text{U}},f_{^{238}\text{U}}, f_{^{239}\text{Pu}}, f_{^{241}\text{Pu}}\}=\{5.5,0.7,3.2,0.6\}\times 10^{-1}$ and $L$ the core-detector baseline, $\mathcal{P}$ is the reactor power and $\varepsilon=205.24$~MeV is the average energy released per fission. The number of $\overline{\nu}_e$ per energy and per fission follows an exponential law
\begin{equation}
    \label{eq:exponential_law}
    \frac{\d N^i_{\overline{\nu}_e}}{\d E_{\overline{\nu}_e}}
    = a_i e^{a_0^i + a_1^i E_{\overline{\nu}_e} + a_2^i E_{\overline{\nu}_e}^2} \ .
\end{equation}
The parameters for each fissile isotope are listed in Tab.~\ref{tab:exp_law_param}. Depending on the detector threshold, the neutrino spectrum below 2 MeV can in 
principle be relevant and may provide a sizable contribution to the total rate. In particular, $\beta$ decays of $^{239}\text{U}$ produced in $^{238}\text{U}(n,\gamma)^{239}\text{U}$ contribute significantly due to their 
large fission fraction. For fluorine, the nucleus that dominates the event rate in $\text{C}_3\text{F}_8$, 
the maximum recoil energy induced by a $2\,$MeV neutrino is 
$E_r^{\mathrm{max}}\sim 0.5~\text{keV}$. Carbon recoils are only slightly larger, reaching at most $E_r^{\mathrm{max}} \sim 0.7\,\text{keV}$. Therefore, for the  detector thresholds considered in this work, this low-energy portion of the 
spectrum does not contribute appreciably to the observable event rate. This  conclusion becomes even stronger for heavier target nuclei, unless detectors  can achieve thresholds well below 1 keV.
\begin{table}[h!]
    \centering
    \setlength{\tabcolsep}{8pt}
    \renewcommand{\arraystretch}{1.5}
    \begin{tabular}{|c||c|c|c|c|}\hline
         \textbf{Param.} & $^{235}\text{U}$ &  $^{239}\text{Pu}$ & $^{238}\text{U}$ & $^{241}\text{Pu}$\\\hline\hline
         $a_{i}\,\mathrm{[MeV^{-1}]}$ & 1.0461 & 1.0527 & 1.0719 & 1.0818\\\hline
         $a_0$ & 0.870 & 0.896 & 0.976 & 0.793\\\hline
         $a_1\,\mathrm{[MeV^{-1}]}$ & -0.160 & -0.239 & -0.162 & -0.080\\\hline
         $a_2\,\mathrm{[MeV^{-2}]}$ & -0.0910 & -0.0981 & -0.0790 & -0.1085\\\hline
    \end{tabular}
    \caption{Parameters for electron antineutrino spectrum   parametrization above $2$ MeV for each fissile isotope~\cite{CONNIE:2019xid}.}
    \label{tab:exp_law_param}
\end{table}

The expected vector and axial event rates are then computed through a convolution of the neutrino flux and the \cevns~differential cross section

\begin{equation}
    \label{eq:diff:dm:events}
    \dfrac{\d N}{\d E_r} = t_{\mathrm{run}} \frac{m_\mathrm{det}}{M_\mathrm{mol}}   \int_{E_\nu^\mathrm{min}}^{E_\nu^\mathrm{max}}\mathrm{d}E_{\nu} \dfrac{\d \Phi_{\nu}}{\d E_\nu} \dfrac{\d \sigma}{\d E_r}\, ,    
\end{equation}
where $t_{\mathrm{run}}$ is the detector data-taking time, $m_\mathrm{det}$ is the detector fiducial mass, and $M_\mathrm{mol}$ is the molar mass of the target material. The lower integration limit depends on the nuclear recoil energy and is expressed as $E_\nu^\text{min}=\sqrt{m_\mathcal{N} E_r/2}$, while $E_\nu^\text{max}$ is taken to be the endpoint of the neutrino energy distribution, i.e., $E_\nu^\text{max}=m_\mu/2$ for the case of ESS and $E_\nu^\text{max}\simeq 10$~MeV for reactor antineutrinos.

\end{widetext}

\end{document}